\documentclass[final,6p,times,twocolumn]{elsarticle}

\usepackage{amssymb}
\usepackage{amsmath}

\usepackage[switch]{lineno}

\journal{Fusion Engineering and Design}

\begin{document}

\begin{frontmatter}



\title{Modelling-driven requirements for Error Field Control Coil application to initial JT-60SA plasmas}


\author[label1]{L. Pigatto} 

\author[label2]{G. Frello}
\author[label3]{L. Garzotti}
\author[label4]{Y.Q. Liu}
\author[label2]{L. Novello}
\author[label5]{M. Takechi}
\author[label1,label6]{E. Tomasina}
\author[label1]{T. Bolzonella}

\affiliation[label1]{organization={Consorzio RFX (CNR, ENEA, INFN, Università di Padova, Acciaierie Venete SpA)}, addressline={Corso Stati Uniti, 4}, city={Padova}, postcode={35127}, country={Italy}}
            
\affiliation[label2]{organization={Fusion for Energy - F4E},
             city={Garching},
             postcode={85748},
             country={Germany}}

\affiliation[label3]{organization={United Kingdom Atomic Energy Authority}, addressline={Culham Campus}, city={Abingdon}, postcode={OX14 3DB},  state={Oxon}, country={UK}}

\affiliation[label4]{organization={General Atomics}, addressline={PO Box 85608}, city={San Diego}, postcode={92186-5608},  state={CA}, country={United States of America}}
 
\affiliation[label5]{organization={National Institutes for Quantum Science and Technology (QST)}, addressline={801-1 Mukoyama}, city={Naka}, postcode={311-0193}, country={Japan}}

\affiliation[label6]{organization={Università degli Studi di Padova}, city={Padova}, postcode={}, country={Italy}}

\begin{abstract}
JT-60SA is a large superconducting tokamak built in Naka, Japan. After the successful achievement of its first MA-class plasma, the installation of several additional sub-systems, including a set of non-axisymmetric Error Field Correction Coils (EFCC), is ongoing. Optimization of future JT-60SA plasma scenarios will critically depend on the correct use of EFCC, including careful fulfillment of system specifications. In addition to that, preparation and risk mitigation of early ITER operations will greatly benefit from the experience gained by early EFCC application to JT-60SA experiments, in particular to optimize error field detection and control strategies. In this work, EFCC application in JT-60SA Initial Research Phase I perspective scenarios is modeled including plasma response. Impact of (Resonant) Magnetic Perturbations on the different plasma scenarios is assessed for both core and pedestal regions by the linear resistive MHD code MARS-F. The dominant core response to EFs is discussed case by case and compared to mode locking thresholds from literature. Typical current/voltage amplitudes and wave-forms are then compared to EFCC specifications in order to assess a safe operational space.
\end{abstract}




\end{frontmatter}




\section{Introduction and background }
\label{sec1}
\begin{figure}
    \centering
    \includegraphics[width=\columnwidth]{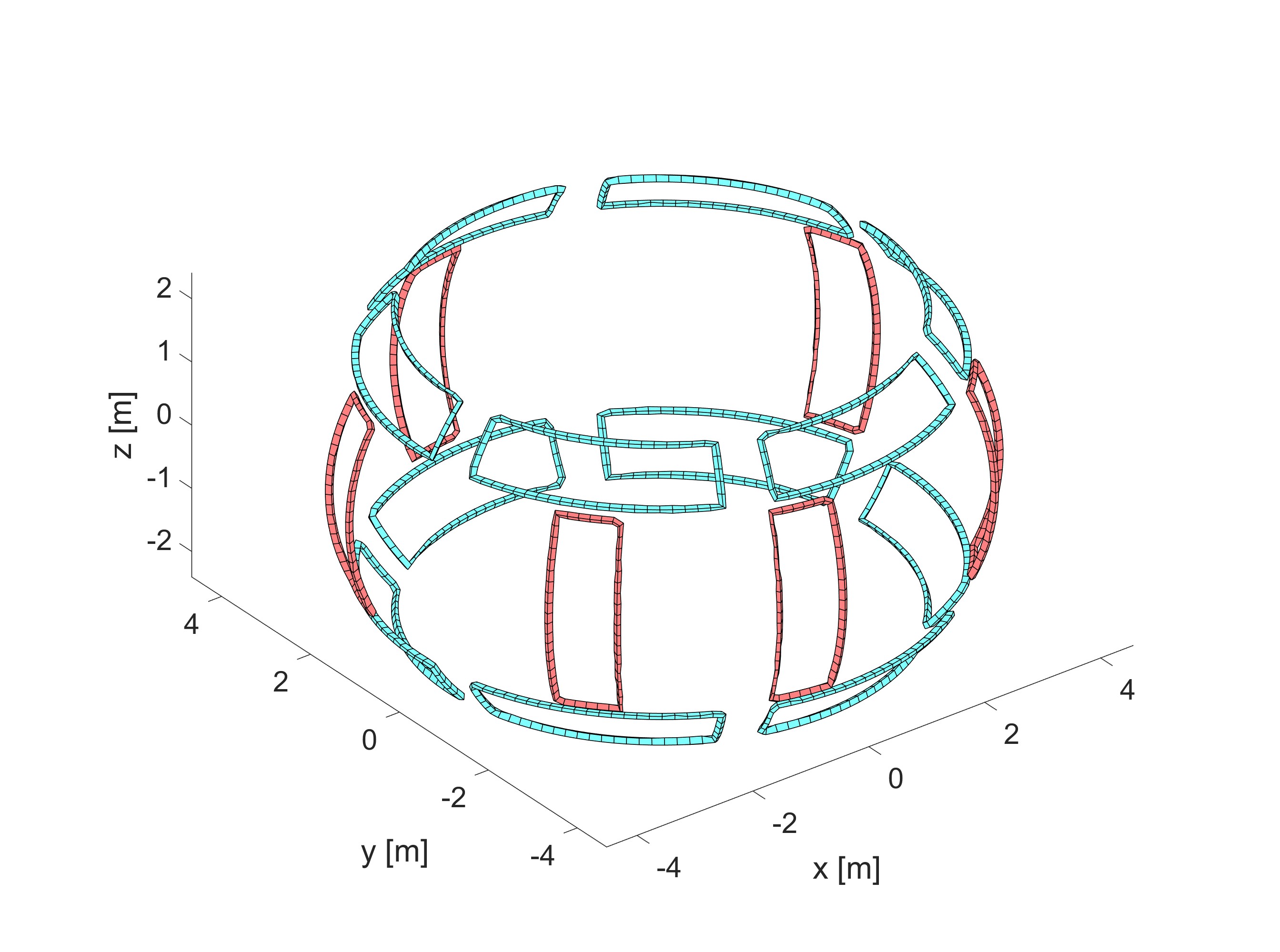}
    \caption{Geometry of JT-60SA Error Field Correction Coils, represented as 3D mesh.}
    \label{fig:efcc}
\end{figure}

The JT-60SA superconducting tokamak \cite{shirai2017recent} was jointly constructed and is jointly funded and exploited under the Broader Approach Agreement between Japan and EURATOM. It started its operations with a successful integrated commissioning carried out in 2023 \cite{davis2024jt}\cite{shirai2023recent}. While this initial phase saw a limited set of installed components and diagnostics \cite{takechi2021vessel}, when the installation of all sub-systems will be completed this device will be well equipped to pursue its main mission of supporting ITER operations and addressing DEMO-relevant physics issues. From the point of view of 3D fields, JT-60SA features two sets of in-vessel coils, often named \textit{saddle coils} due to their shape. The first set is dedicated to feedback control of Resistive Wall Modes \cite{takechi2023design}, Magneto-Hydro-Dynamic instabilities which can develop in high performance plasmas and their stabilization is thus important for the mission of JT-60SA towards steady-state high $\beta$ operation \cite{pigatto2019resistive}. A second system of saddle coils, named Error Field Correction Coils (EFCC) and carrying a peak current of 52.5 kAt, will be mounted on the vacuum vessel inner outboard and organized in 3 (Upper, Mid-plane, Lower) sets of 6 coils each. The 18 in-vessel EFCC, shown in Figure \ref{fig:efcc}, will have the main target of correcting residual non-axisymmetric magnetic field components generated e.g. by misalignment of the equilibrium coils, i.e. Error Fields (EF), and of controlling Edge Localized Modes (ELMs) by ergodization of the field at the edge.

\section{Error Field and plasma response modeling}
\label{sec2}
Progress in EF studies over the last decade has led to understanding that not only the vacuum non-axisymmetric components of the equilibrium need to be corrected for smooth operations, but the response of a given plasma to such components is of paramount importance. Eventually plasma response dominates over the applied vacuum field and what should be reduced is the kink-resonant error field component, i.e. the one associated with the largest plasma kink-like response. A \textit{dominant external field} concept can be used to define the distribution of external perturbation that drives the maximum core plasma response. The convolutional integral between the external and plasma response field distributions can be computed to quantify how the two overlap. Any EF is just as dangerous as much as it overlaps with this dominant external field \cite{park2008error}\cite{park2011error}. To prevent EF penetration, the component that overlaps with the dominant external field must be reduced below a threshold. Empirical scaling laws exist for this threshold, such as those reported in \cite{logan2020empirical}.

Response to n=1 perturbations is of particular importance for the impact EFs can have on a given plasma. Thus we studied the coupling of n=1 external fields, with varying poloidal (m) harmonics, to a JT-60SA relevant plasma, with toroidal field $B_t = 1.8 T$ and plasma current $I_p = 2.1 MA$. This choice reflects the main target scenario for upcoming experiments to be carried out during the so-called \textit{Operational Phase 2}. Plasma response is studied with the MARS-F linear resistive MHD code \cite{liu2000} \cite{liu2010}. A workflow for the calculation of linear plasma response models. (PRM) has been developed, building on the Equivalent Surface Current (ESC) method described in \cite{liu2014ESC}. These models can be applied for the calculation of EF correction recipes that include plasma response (including the effect of resistivity and toroidal flow captured by MARS-F), as recently tested on AUG \cite{igochine2023plasma}. In this workflow, we select a single toroidal mode number ($n=1$) and a range of poloidal harmonics. Each poloidal harmonic is used as external perturbation on a selected coupling surface, external to the plasma, and plasma response is calculated with the aforementioned ESC method. We would like to highlight that these perturbations are implemented here as surface currents in a vacuum region of the computational domain, not as boundary conditions as in similar applications \cite{portone2008linearly}. The result for all poloidal harmonics is a database that can be used to model, via linear combinations, plasma response to any external field represented with the selected poloidal spectrum. Such a database has been analyzed with Singular Value Decomposition to find the dominant modes, i.e. the components to which the target plasma is most sensitive.

With the calculated models, EF correction predictions can be made based on the overlap field criterion \cite{park2008error}, which aims at correcting the components of the penetrated vacuum field which better overlap the plasma response pattern. In the typical expression of the overlap field $b_{ovlp}^*(n) = \sum_m V_{mn} \times b_{mn}$, the modes from SVD are labelled $V_{mn}$ and the external vacuum field is $b_{mn}$. 
As described in the following sections, these SVD modes will be used to calculate the overlap of a given EF with plasma response in a given scenarios as well as the coefficients stating how efficiently a given coil set is driving plasma response (i.e. in correcting the EF).
The calculation of EF control currents requires then unit fields from EFCC and proxy EFs from the possible sources (e.g. PF, CS, TF coils). Future work shall investigate different sources of EF in JT-60SA, following Monte-Carlo studies such as \cite{matsunaga2015vessel}, and predict possible correction recipes including resistive physics and toroidal flow. For ELM control studies instead, linear modelling suggests that the required coil currents are well within the system’s capabilities.

\section{Development of dynamic simulation tool}
\label{sec3}

A Matlab Simulink model has been developed for calculating reference EFCC currents and modeling the actual load currents. The blocks used in this model are described hereafter, these include: calculation of the overlap Error Field, calculation of the coil current reference, model of the power supply and coil dummy load. 

\subsection{Calculation of overlap EF}
\label{sec31}

\begin{figure*}[h!]
    \centering
    \includegraphics[width=\textwidth]{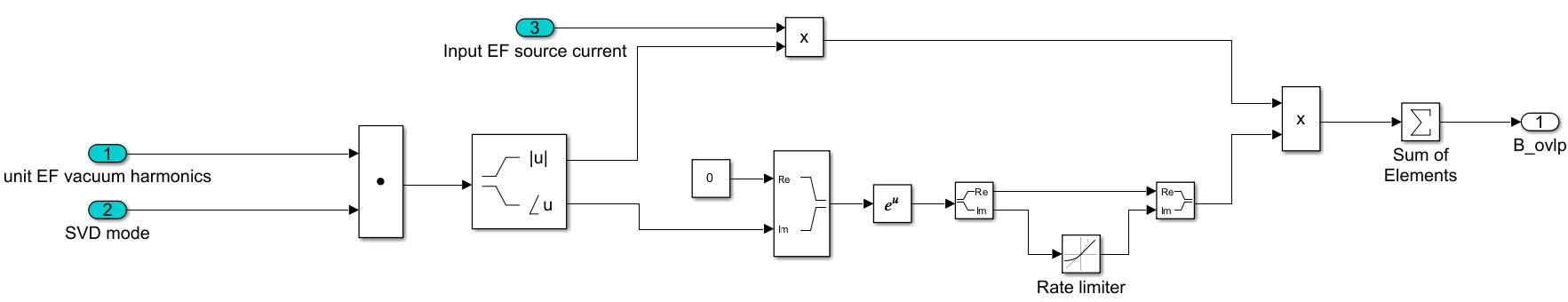}
    \caption{Simulink block for calculation of overlap field.}
    \label{fig:simulink_bovlp}
\end{figure*}
Calculation of the overlap field follows the description in Section \ref{sec2}, with the additional feature of scaling the overlap EF magnitude with the current flowing in the EF source. In the present model the EF is assumed to be generated by an $n=1$ non-axisymmetric magnetic field component from the central solenoid (CS) segments. From the machine point of view such a component could be generated by misalignment or deformation of the coils. We highlight that this assumption on the EF source is used as a worst case scenario for proof-of-principle and does not correspond to a realistic EF, which will be investigated in future work. The Simulink block shown in Figure \ref{fig:simulink_bovlp} represents the following expression:  
\begin{equation}
B_{OVLP}(t) = |b_{ovlp}^{*}|I^{EF}e^{i(\angle b_{ovlp}^{*})}  
\end{equation}
where $I^{EF}$ is the EF source current and $b_{ovlp}^{*}$ is calculated using the first SVD mode right-singular vector and the unit vacuum EF poloidal spectrum from each source.

\begin{figure}[h!]
    \centering
    \includegraphics[width=\columnwidth]{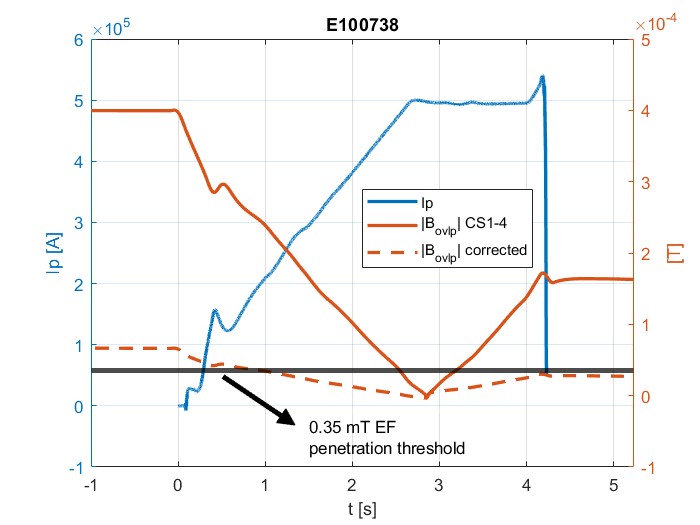}
    \caption{Right y-axis: total overlap error field for CS1-2 (solid) and corrected overlap field (dashed). Left y-axis: plasma current assuming E100738 as possible target shot.}
    \label{fig:bovlp}
\end{figure}
For developing this workflow with the aforementioned assumptions on the EF source, coil currents have been taken from experiments carried out during Integrated Commissioning \cite{davis2024jt}. These experiments are not necessarily corresponding to the scenario used for plasma response. While this inconsistency will be addressed in future work, plasma response patterns are found to be similar for L-mode scenarios, this approach has been therefore followed for development. Figure \ref{fig:bovlp} shows the plasma current and overlap field time traces for shot $E100738$. $B_{OVLP}(t)$ varies in time following the CS current evolution, we note that in some phases the threshold calculated for this scenario (using the $n=1$ scaling law in \cite{logan2020empirical}) is exceeded ($B_{OVLP} \simeq 0.35 mT$). For the case reported in Figure \ref{fig:bovlp} this happens during transients, where the applicability of EF correction has not been thoroughly investigated and is out-of-scope for the present work. This shot has been chosen for proof-of-principle because at this level of plasma current some first EF identification and correction experiments could be envisaged in the future. The residual EF, including correction, is also reported and is found to be lower than the aforementioned penetration threshold in the relevant plasma flat-top phase.

\subsection{Calculation of EFCC current reference}
\label{sec32}

Using the overlap field calculated in the previous block, references for coil current can be obtained with the following:
\begin{align}
I_i &= I_0 \cos(n \phi_k - \phi) \\
k &= 1, ..., N_{\text{coils}} \\
I_0 &= -\frac{|B_{OVLP}(t)|}{b_0} \\
\phi &= \alpha - \phi_0
\end{align}
where $\phi_k$ is the coil center angle, ($b_0$, $\phi_0$) are the overlap coefficients for each coil row (representing the efficiency in driving correction field), $\alpha$ is the phase of the overlap EF. The above equations have been translated in a Simulink block where the coil geometry and efficiency values ($b_0$, $\phi_0$) are taken as input, together with the calculated $B_{OVLP}(t)$.

\subsection{Modeling EFCC power supply}
\label{sec33}

A single modeling block is implemented for calculating the reference voltage, power supply modeling and coil dummy load. From the previously calculated coil current references, the voltage is obtained with a P-I controller (with gains $K_P=0.5$, $K_I=0.25$ respectively). The current reference is limited within $\pm 1500 A$ with a maximum rate of $70 kA/s$. The reference voltage is limited to the range $\pm 100 V$. The EFCC inverter is modeled with $350 \mu s$ delay and a proportional gain of $4.5$, implying a maximum output voltage of $450 V$.


\subsection{Discussion}
\label{sec34}
\begin{figure}
    \centering
    \includegraphics[width=\columnwidth]{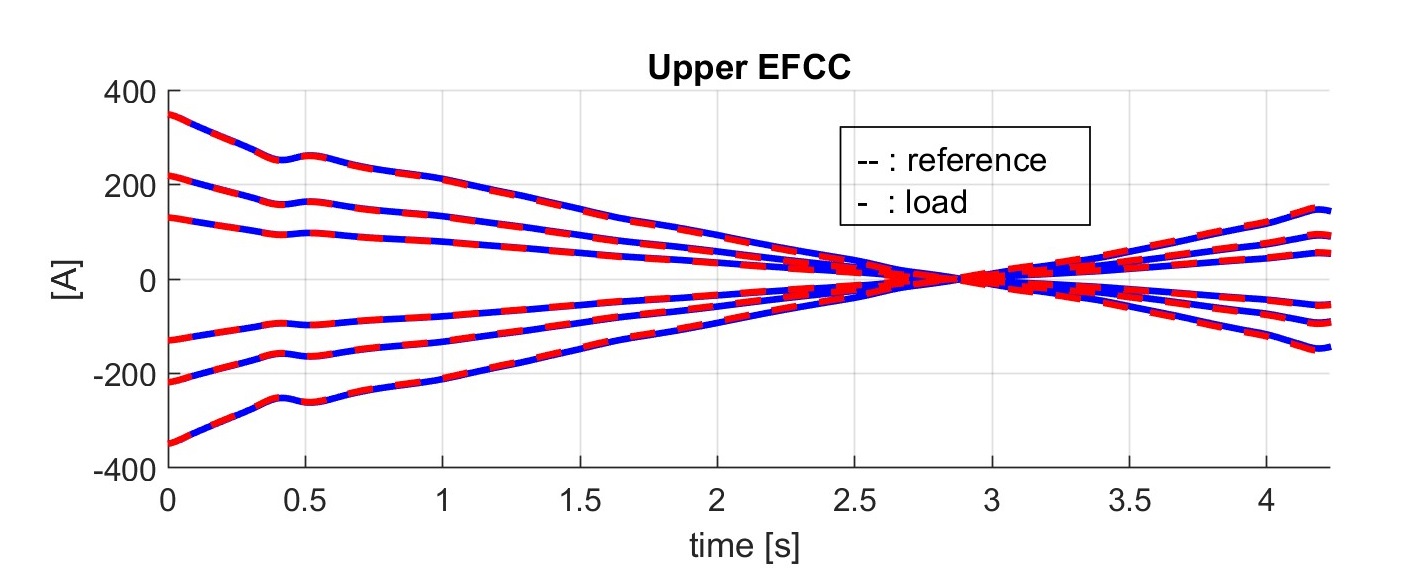}
    \caption{Time evolution of currents in upper EFCC. Reference wave forms are shown in red dashed lines, load currents as blue solid lines.}
    \label{fig:loadVSREF}
\end{figure}

\begin{figure}
    \centering
    \includegraphics[width=\columnwidth]{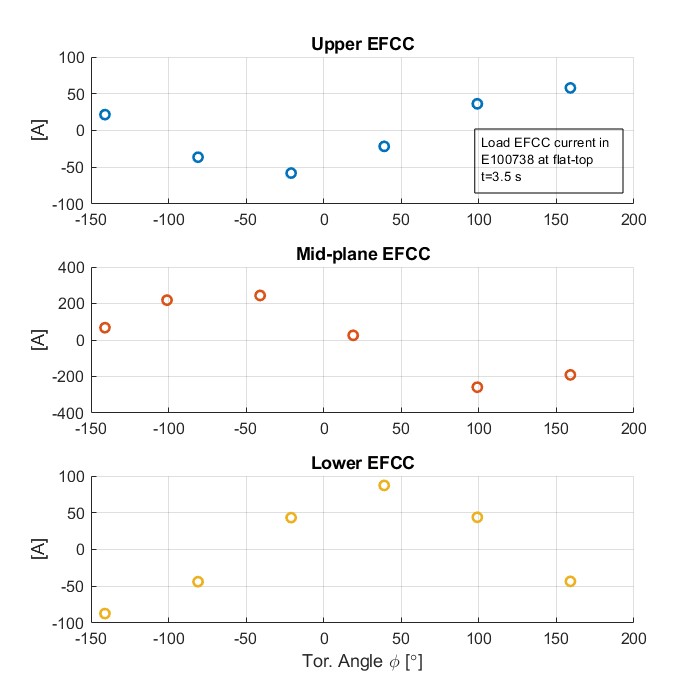}
    \caption{Coil currents in the Upper, Mid-plane, and Lower EFCC (independently operated) at t=3.5s during flat-top phase.}
    \label{fig:currFlattop}
\end{figure}

As depicted in Figure \ref{fig:loadVSREF}, the load currents in the upper coils closely track the desired reference currents during the selected discharge (E100738). This tracking is achieved while considering the assumed EF spectrum generated by the central solenoid. Figure \ref{fig:currFlattop} further illustrates the sinusoidal current pattern in each coil across all rows, captured at a fixed time during the plasma current flat-top. The sinusoidal nature of the currents is expected, as it is a characteristic of the EF correction scheme employed. The three coil rows (upper, mid-plane, lower) are assumed to operate independently, each aiming to fully compensate the overlapping EF. This assumption implies that each coil row can be controlled individually to achieve the desired EF correction. It is noteworthy that the mid-plane coils require higher currents compared to the upper and lower coils. This discrepancy can be attributed to several factors: specific plasma conditions can influence the response and thus the required EF correction, the assumed EF spectrum and topology in the mid-plane region can impact the required coil currents, the narrower toroidal geometry of the mid-plane coils compared to the upper and lower coils can lead to weaker coupling to n=1 fields. This reduced coupling efficiency necessitates higher currents to achieve the same level of EF correction.

\section{Conclusion}
This work investigated the application of Error Field Correction Coils (EFCC) in the JT-60SA tokamak. We modeled the plasma response to resonant magnetic perturbations and assessed their impact on core and pedestal regions. The dominant core response was analyzed for selected plasma scenarios and compared to mode locking thresholds. Our findings indicate that the EFCC system can effectively mitigate error fields and control plasma instabilities. We have developed a dynamic simulation tool to predict reference EFCC currents and model the actual load currents. This tool can be used for optimizing EFCC operation and ensuring the successful implementation of error field correction strategies. Future research efforts will focus on refining the EF model to better capture the actual plasma conditions. Additionally, investigations into optimized EF correction strategies for multiple coil rows, considering their interdependence and synergistic effects, will be pursued. A detailed paper outlining the workflow for plasma response calculations is currently in preparation.

\section*{Acknowledgments}
The authors would like to thank Dr. N. Ferron and Prof. L. Piron for the useful discussions of error field control and model implementation. This work has been carried out within the framework of the EUROfusion Consortium, funded by the European Union via the Euratom Research and Training Programme (Grant Agreement No 101052200 — EUROfusion). Views and opinions expressed are however those of the author(s) only and do not necessarily reflect those of the European Union or the European Commission. Neither the European Union nor the European Commission can be held responsible for them. 
This material is based upon work partly supported by the U.S. Department of Energy, Office of Science, Office of Fusion Energy Sciences, under Award DE-FG02-95ER54309. Part of the data analysis was performed using the OMFIT integrated modeling framework \cite{meneghini2015}.


\bibliographystyle{elsarticle-num} 
\bibliography{soft24_EF_JT60SA.bib}

\end{document}